\documentclass[prd,twocolumn,showpacs,preprintnumbers]{revtex4}
\usepackage{hyperref,amssymb,amsmath,mathrsfs,bm,graphicx}
\usepackage{subfigure}
\newcommand{\be}{\begin{equation}}
\newcommand{\ee}{\end{equation}}

\def\ltsima{$\; \buildrel < \over \sim \;$}
\def\lsim{\lower.5ex\hbox{\ltsima}}
\def\gtsima{$\; \buildrel > \over \sim \;$}
\def\gsim{\lower.5ex\hbox{\gtsima}}

\begin{document}
\title{Shear Viscosity in the Postquasistatic Approximation}
\author{C. Peralta}
\affiliation{Deutscher Wetterdienst, Frankfurter Str. 135, 63067 Offenbach, Germany}
\altaffiliation[Also at: ]{School of Physics, University of Melbourne, Parkville, VIC 3010, Australia}
\author{L. Rosales}
\affiliation{Laboratorio de F\'\i sica Computacional,
Universidad Experimental Polit\'ecnica ``Antonio Jos\'e de
Sucre'', Puerto Ordaz, Venezuela}
\author{B. Rodr\'\i guez--Mueller}
\affiliation{Computational Science Research Center, College of Sciences,
San Diego State University, San Diego, California, USA}
\author{ W. Barreto}
\affiliation{Centro de F\'\i sica Fundamental, Facultad de Ciencias, Universidad de Los Andes, M\'erida, Venezuela}
\date{\today}
\begin{abstract}
We apply the postquasistatic approximation, 
an iterative method for the evolution of self--gravitating spheres of matter,
to study the evolution of anisotropic non--adiabatic radiating and dissipative 
distributions in General Relativity.  
Dissipation is described by viscosity and free--streaming radiation,
assuming an equation of state to model anisotropy induced
by the shear viscosity. We match the interior solution, in non--comoving coordinates, with
the Vaidya exterior solution. Two simple models are presented, based
on the Schwarzschild and Tolman VI solutions, in the non--adiabatic
and adiabatic limit. In both cases the eventual collapse or expansion
of the distribution is mainly controlled by the anisotropy induced by the viscosity.
\end{abstract}

\maketitle

\section{Introduction}

In order to study astrophysical fluid dynamics one can get complicated models
incorporating realistic transport mechanisms and equations of state.
The simplest case with mass, spherical symmetry, despite its simplicity,
still remains an interesting problem in numerical relativity, specially
when including dissipation.  Dissipation due to the emission of massless
particles (photons and/or neutrinos) is a characteristic process
in the evolution of massive stars. It seems that the only plausible mechanism
to carry away the bulk of the binding energy of the collapsing star,
leading to a black hole or neutron star, is neutrino emission \cite{ks79}.
Viscosity may be important in the neutrino trapping
during gravitational collapse \cite{arnett77,brown82,bethe82}, which is expected to occur when
the central density is of the order $10^{11}$--$10^{12}$g cm$^{-3}$. Although
the mean free path of the neutrinos is much greater than for others particles,
the radiative Reynolds number of the trapped neutrinos is nevertheless small
at high density \cite{mm84}, rendering viscous the core fluid \cite{k78},
\cite{ks79}.

Numerical Relativity is expected to keep its power to solve problems and generate new,
interesting physics, when dissipative distributions of matter are considered.
In fact, numerical methods in General Relativity have been proven to be extremely
valuable for the investigation of strong field scenarios (see \cite{lehner}
and references therein). For instance, these methods and frameworks have (i)
revealed unexpected phenomena \cite{choptuik}, (ii) enabled the simulation of
binary black holes (neutron stars) \cite{pretorius,bgr08} and
(iii) allowed the development of relativistic hydrodynamic solvers \cite{font},
among other achievements. Currently, the main limitation for Numerical Relativity
is the computational demand for 3D evolution \cite{winicour}.
The addition of a test-bed for studying dissipation mechanisms and other
transport processes in order to later incorporate them into a more sophisticated
numerical framework (ADM or characteristic) is a necessity.

In this paper, we study 
a selfgravitating spherical distribution of matter containing a
dissipative fluid.  We follow the method proposed in
\cite{brm2001}, which introduces a set
of conveniently defined ``effective'' variables (effective pressure
and energy density), where their radial
dependence is chosen on heuristic grounds. In essence this is equivalent
to going one step further from the quasistatic regime, and
the method has been named the
postquasistatic approximation (PQSA) after \cite{hbds2001}.
The essence of the PQSA was first proposed in \cite{hjr80} using radiative
Bondi coordinates and it has been extensively used by Herrera and collaborators \cite{hn90FCP,bn91,bhn91,bhn92,abn94,hmnp94,m96,bds96}.
By quasistatic approximation we mean
that the effective variables coincide with the corresponding physical
variables (pressure and energy density).
However, in Bondi coordinates the notion of quasistatic
approximation is not evident: the system
goes directly from static to postquasistatic evolution. In an adiabatic
and slow evolution we can catch--up that phase, clearly seen in non--comoving
coordinates. This can be achieved using Schwarzschild coordinates \cite{hbds2001}.
Here we study radiating viscous fluid spheres in the streaming
out limit with the
PQSA approach which allow us to departure from equilibrium in non--comoving coordinates.
These systems have been studied using the method described in \cite{hjr80}
for the radiative shear viscosity
problem and its effect on the relativistic gravitational collapse 
\cite{hjb89,br92,b93,bc95,bor98}. 
We do not consider temperature
profiles to determine which processes can take place during the collapse.
For that purpose, transport equations in the relaxation time
approximation have been proposed to avoid
pathological behaviors (see for instance \cite{m96} and references therein). These issues
will be considered in a future investigation. 
In order to develop a numerical solver which
incorporates in a realistic way dissipation following
the M\"uller-Israel-Stewart theory \cite{m67,i76,is76,is79}
it is first necessary to know, to zeroth level of approximation, viscosity
profiles like the ones presented in this investigation.
The physical consequences of considering dissipation by means
of an appropriate causal procedure have been stated analytically
in several papers by Herrera and collaborators (see for example 
\cite{hdhmm97,hm98,hdmost04,hdb06}).

To the best of our knowledge, no
author has undertaken in practice the dissipative matter problem in
numerical relativity. 
Our purpose here is to show how viscosity processes can be considered as
anisotropy and how the PQSA works in this context.
Our results partially confirm previous investigations \cite{br92,b93}.
The novelty here is in the use of the PQSA to study dissipative
scenarios. The results indicate that an observer using radiation coordinates
does not "see'' some details when shear viscosity is considered.
The final goal is to eventually study the same problem using the 
M\"uller-Israel-Stewart theory for
dissipative system, which is highly nontrivial in spherical symmetry.

In standard numerical relativity, in order to deal with matter both in
ADM 3+1 \cite{nc00} and in the characteristic
formulations \cite{font}, Bondian observers have been
used implicitly. This has been noted 
recently, and the method has been proposed as a test bed in numerical relativity
\cite{b09}. The systematic use of local Minkowskian and comoving observers in the PQSA,
named Bondians, was used to reveal a central equation of state
in adiabatic scenarios \cite{bcb09}, and to couple matter with radiation \cite{bcb10}. 
Since Bondian observers are a fundamental part of the PQSA
and all its applications in the characteristic formulation we
are currently trying to transfer all the experience gained
using this approach to include more realistic effects in the
dynamics of the fluid using the ADM 3+1 formulation, the most
popular method in numerical relativity.
Besides introducing a more realistic time scale in the problem with matter
the intention is to promote the PQSA (and any of its applications) as 
a test bed in the ADM 3+1 and characteristic approaches.

The plan for this paper is as follows: In Section \S\ref{sec:FEM} we present the field
equations and matching conditions at the surface of the distribution. We explain the
PQSA and write a set of surface equations, in Section \S\ref{sec:pqsa}. In Section \S\ref{sec:examples}
we illustrate the method presenting four simple models based on
the Schwarzschild and Tolman VI interior solutions. Finally, we discuss the
results in Section \S\ref{sec:conc}.

\section{Field equations for Bondian frames and matching}
\label{sec:FEM}
~~~~To write the Einstein field equations we use the line element in Schwarzschild--like coordinates
\begin{equation}
ds^2=e^\nu dt^2-e^\lambda dr^2-r^2\left( d\theta ^2+\sin
{}^2\theta d\phi ^2\right), \label{eq:metric} 
\end{equation}
where $\nu = \nu(t,r)$ and $\lambda = \lambda(t,r)$, with
 $(t,r,\theta,\phi)\equiv(0,1,2,3)$.

In order to get physical input we introduce the
 Minkowski coordinates $(\tau,x,y,z)$ by \cite{b64}
\begin{equation}
d\tau=e^{\nu /2}dt,\,  
dx=e^{\lambda /2}dr,\,  
dy=rd\theta,\, 
dz=r \sin \theta d\phi,\label{eq:local}
\end{equation}
In these expressions $\nu$ and $\lambda$ are constants, because they
 have only local values. 

Next we assume that, for an observer moving relative to these coordinates
 with velocity $\omega$ in the radial ($x$) direction, the space contains
\begin{itemize}
\item a viscous fluid of density $\rho$, pressure $\hat p$, effective
 bulk pressure $p_\zeta$ and effective shear pressure $p_\eta$, and
\item unpolarized radiation of energy density $\hat \epsilon$.
\end{itemize}

For this moving observer, the covariant energy tensor in Minkowski
 coordinates is thus
 
\begin{equation}
\left(
\begin{array}{cccc}
\rho+\hat \epsilon & -\hat \epsilon & 0 & 0 \\
-\hat \epsilon & \hat p +\hat \epsilon - p_\zeta - 2 p_\eta& 0 & 0 \\
0 & 0 & \hat p - p_\zeta + p_\eta & 0 \\
0 & 0 & 0 & \hat p - p_\zeta + p_\eta
\end{array}
\right).
\end{equation}

Note that from (\ref{eq:local}) the velocity of matter in Schwarzschild
 coordinates is
\begin{equation}
\frac{dr}{dt} = \omega e^{(\nu-\lambda)/2}. \label{eq:velocity}
\end{equation}

Making a Lorentz boost and defining  
 $\bar p \equiv \hat p - p_\zeta$,
 $p_r \equiv \bar p - 2 p_\eta$,
 $p_t \equiv \bar p + p_\eta$ and
 $\epsilon \equiv \hat \epsilon(1+\omega)/(1-\omega)$ we write 
 the field equations in relativistic units ($G=c=1$) as follows: 
\begin{equation}
\tilde\rho =
\frac{1}{8\pi r}\left[\frac{1}{r} - 
e^{-\lambda}\left(\frac 1{r}-\lambda_{,r}\right)\right], \label{eq:ee1}
\end{equation}

\begin{equation}
\tilde p =
\frac{1}{8\pi r}\left[
e^{-\lambda}\left(\frac 1{r}+\nu_{,r}\right) - \frac{1}{r}\right], \label{eq:ee2}
\end{equation}

\begin{eqnarray}
p_t = &&\frac{1}{32\pi} \{ e^{-\lambda}[ 2\nu_{,rr}+\nu_{,r}^2
-\lambda_{,r}\nu_{,r} + \frac{2}{r}
(\nu_{,r}-\lambda_{,r}) ] \nonumber \\ \nonumber \\
&-&e^{-\nu}[ 2\lambda _{,tt}+\lambda_{,t}(\lambda_{,t}-\nu_{,t}) ] \}, \label{eq:ee3}
\end{eqnarray}

\begin{equation}
S = 
-\frac{\lambda_{,t}}{8\pi r}e^{-\frac 12(\nu+\lambda)}, \label{eq:ee4}
\end{equation}
where the comma (,) represents partial differentiation with 
 respect to the indicated
 coordinate and the effective variables $\tilde\rho$, $S$, known as conservation variables  as well, and $\tilde p$, the flux variable,
 \begin{equation}
\tilde\rho= \frac{\rho + p_r \omega^2}{1-\omega ^2} + \epsilon, \label{eq:ev1}
\end{equation}
\begin{equation}
S=(\rho + p_r)\frac{\omega}{1-\omega^2} + \epsilon
\end{equation}

and
\begin{equation}
\tilde p = \frac{p_r + \rho \omega^2}{1-\omega ^2} + \epsilon. \label{eq:ev2}
\end{equation}

Equations (\ref{eq:ee1})--(\ref{eq:ee4}) are formally the same as for
 an anisotropic fluid
 in the streaming out approximation \cite{b93}.

At this point, for the sake of completeness, we write the effective
 viscous pressures in terms of
 the bulk viscosity $\zeta$, the volume expansion $\Theta$, the shear
 viscosity $\eta$ and the scalar shear $\sigma$ 
\begin{equation}
p_\zeta = \zeta \Theta,
\end{equation}
\begin{equation}
p_\eta = \frac{2}{\sqrt{3}} \eta \sigma,
\end{equation}
where
\begin{eqnarray}
\Theta &=& \frac{1}{(1-\omega^2)^{1/2}}\left[e^{-\nu/2}\left(\frac{\lambda_{,t}}{2}
+  \frac{\omega\omega_{,t}}{1-\omega^2}\right)\right.\nonumber\\ \nonumber\\
&&\left.+e^{-\lambda /2}\left(\frac{\nu_{,r}}{2}\omega+\frac{1+\omega^2}{1-\omega^2}\omega_{,r}+\frac{2\omega}
{r} \right) \right] \label{eq:expansion}
\end{eqnarray}
and
\begin{equation}
\sigma =\pm\sqrt{3}\left( \frac \Theta 3-\frac{e^{-\lambda /2}}r\frac
\omega {\sqrt{1-\omega ^2}}\right). \label{eq:shear}
\end{equation}

We have four field equations for five physical variables ($\rho$, $p_r$,
 $\epsilon$, $\omega$ and $p_t$) and two geometrical
 variables ($\nu$ and $\lambda$). Obviously, we require additional
 assumptions to handle the problem consistently. However, we
 discuss first the matching with an exterior solution and the surface
 equations that govern the dynamics.

~~~~We describe the exterior space--time by the Vaidya metric 
\begin{equation}
ds^2=\left( 1-\frac{2{\cal M}(u)}R\right) du^2+2dudR-R^2\left( d\theta
^2+\sin^2\theta d\phi^2 \right),
\end{equation}
where $u$ is a time--like coordinate so that $u=$ constant represents,
 asymptotically,  null
 cones open to the future and R is a null coordinate ($g_{RR}=0$). 
 The relationship at the surface between the coordinates ($t$,$r$,$\theta$,$\phi$)
 and ($u$,$R$,$\theta$,$\phi$) is
\begin{equation}
u=t-r-2{\cal M}\ln \left( \frac r{2{\cal M}}-1\right), R=r.
\end{equation}

The exterior and interior solutions are separated by the surface $r=a(t)$.
 In order to match both regions on this surface we use the Darmois junction
 conditions. Demanding the continuity of the first fundamental form, we obtain
\begin{equation}
e^{-\lambda_a}=1-\frac{2{\cal M}}{R_a} \label{eq:ffa}
\end{equation}
and 
\begin{equation}
\nu_a = -\lambda_a. \label{eq:ffb}
\end{equation}
From now on the subscript $a$ indicates that the quantity is evaluated 
 at the surface.
 Now, instead of writing the junction conditions as usual, we demand the continuity
 of the first fundamental form and the continuity of the 
 independent components of the energy--momentum flow \cite{hdp97}. This last condition
 guarantees the absence of singular behaviors on the
 surface. It is easy to check that
\begin{equation}
\hat p_a = p_{\zeta _a} + 2 p_{\eta _a}, \label{eq:boundary}
\end{equation}
which expresses the discontinuity of the radial pressure in the presence of
 viscous processes.

Before proceeding with the description of the method 
it is convenient to rewrite some
equations and introduce one equation of state.

Defining the mass function as
\begin{equation}
e^{-\lambda}=1-2m/r, \label{eq:mass}
\end{equation}
and substituting (\ref{eq:mass}) into (\ref{eq:ee1}) and (\ref{eq:ee4})
 we obtain, after some arrangements, 
\begin{equation}
\frac{dm}{dt}=-4\pi r^2\left[\frac{dr}{dt}p_r
+\epsilon (1-\omega )(1-2m/r)^{1/2}e^{\nu /2} \right]. \label{eq:energy}
\end{equation}
This equation, known as the momentum constraint in the ADM 3+1 formulation,
expresses the power across any moving spherical shell.

Equation (\ref{eq:ee3}) can be written as $T_{1;\mu }^\mu=0$ or
equivalently, after a lengthly calculation
\begin{eqnarray} 
&&\tilde p_{,r} + \frac{(\tilde\rho + \tilde p)(4\pi r^3\tilde p + m)}{r(r-2m)}+\frac{2}{r}(\tilde p-p_t)=\nonumber\\
&&\frac{e^{-\nu}}{4\pi r(r-2m)}\left( m_{,tt} +\frac{3m_{,t}^2}{r-2m}-
\frac{m_{,t}\nu_{,t}}{2}\right).  \label{eq:TOV}
\end{eqnarray}
This last equation corresponds to a generalization
of the hydrostatic support equation, that is, the Tolman--Oppenheimer--Volkoff (TOV) equation.
 It can be shown that equation (\ref{eq:TOV}) is equivalent to the equation
of motion for the fluid in conservative form in the standard ADM 3+1 formulation \cite{b09}. 
Equation (\ref{eq:TOV}) leads to the third equation at the surface (see next section); up to this point
is completely general within spherical symmetry.

To close this section we have to mention that we assume the following equation of state \cite{b93} for
non--adiabatic modeling
\cite{chew82}
\begin{equation}
p_t - p_r = \frac{C (\tilde{p} + \tilde{\rho})(4 \pi r^{3} \tilde{p} +
m)}{(r-2m)}
\label{eq:EDE}
\end{equation}
where $C$ is a constant.

\begin{figure}[htbp!]
\begin{center}
\scalebox{0.4}{\includegraphics[angle=270]{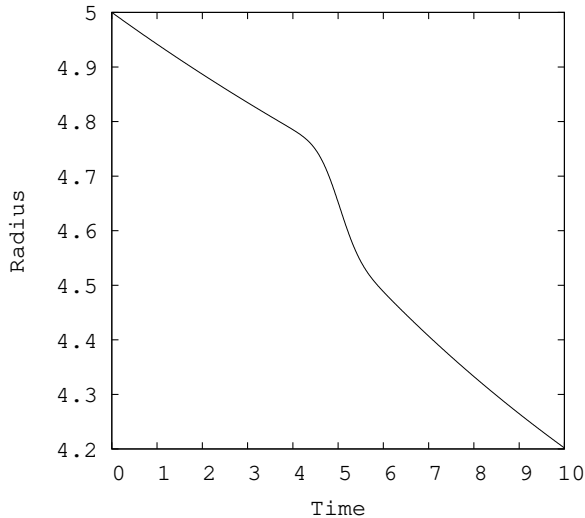}}
\caption{Evolution of the radius $A(t)$ for the Schwarzschild--like model I. 
The initial conditions are $A(0) = 5.0$, $F(0)=0.6$, $\Omega(0)=-0.1$.}
\end{center}
\label{fig:rsch}
\end{figure}
\begin{figure}[htbp!]
\begin{center}
\scalebox{0.4}{\includegraphics[angle=270]{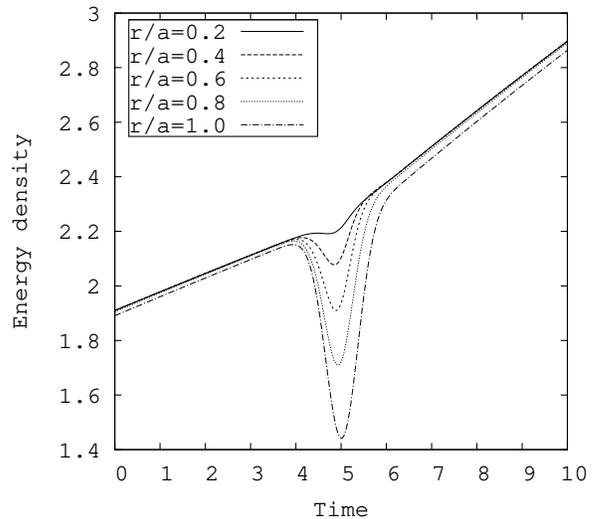}}
\caption{Evolution of the energy density ${\rho}$ (multiplied by $10^3$) for the Schwarzschild--like model I.
The initial conditions are $A(0) = 5.0$, $F(0)=0.6$, $\Omega(0)=-0.1$, with $h=0.99$.}
\end{center}
\label{fig:dsch}
\end{figure}
\begin{figure}[htbp!]
\begin{center}
\scalebox{0.4}{\includegraphics[angle=270]{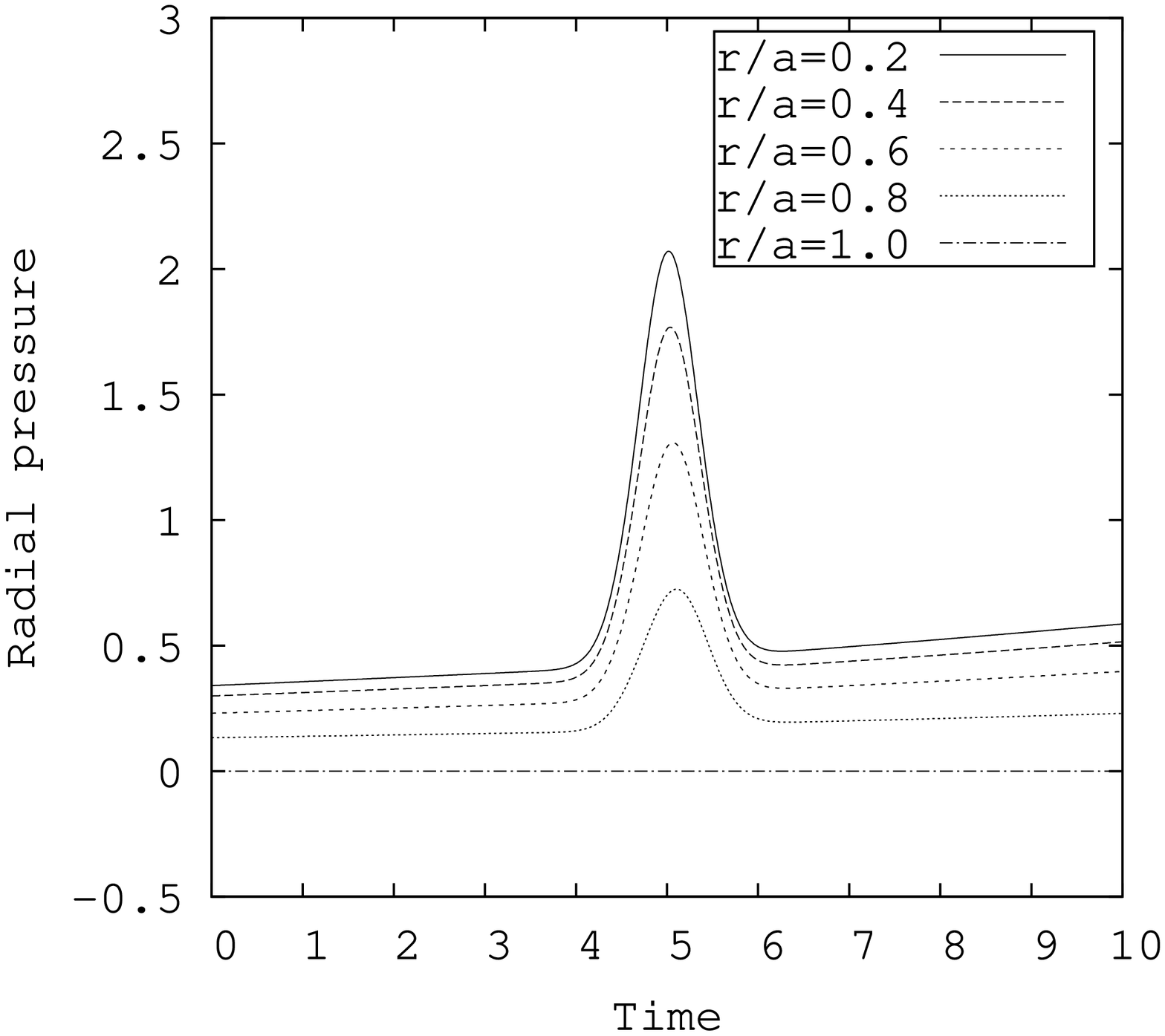}}
\caption{Evolution of the radial pressure $p_r$ (multiplied by $10^3$) for the Schwarzschild--like model I. 
The initial conditions are $A(0) = 5.0$, $F(0)=0.6$, $\Omega(0)=-0.1$, with $h=0.99$.}
\end{center}
\label{fig:psch}
\end{figure}
\begin{figure}[htbp!]
\begin{center}
\scalebox{0.4}{\includegraphics[angle=270]{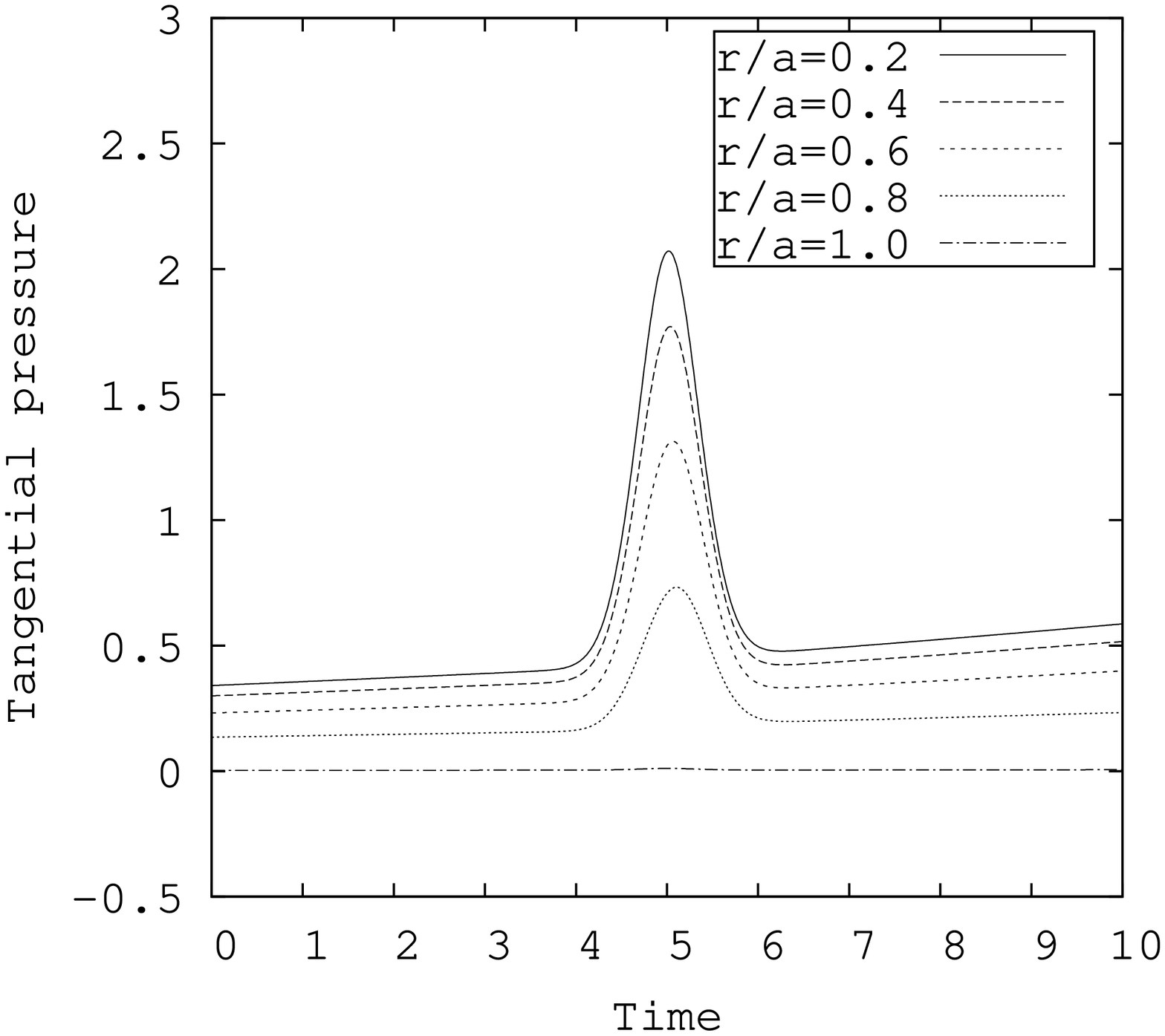}}
\caption{Evolution of the tangential pressure $p_t$ (multiplied by $10^3$) for the Schwarzschild--like model I. 
The initial conditions are $A(0) = 5.0$, $F(0)=0.6$, $\Omega(0)=-0.1$, with $h=0.99$.}
\end{center}
\label{fig:ptsch}
\end{figure}
\begin{figure}[htbp!]
\begin{center}
\scalebox{0.4}{\includegraphics[angle=270]{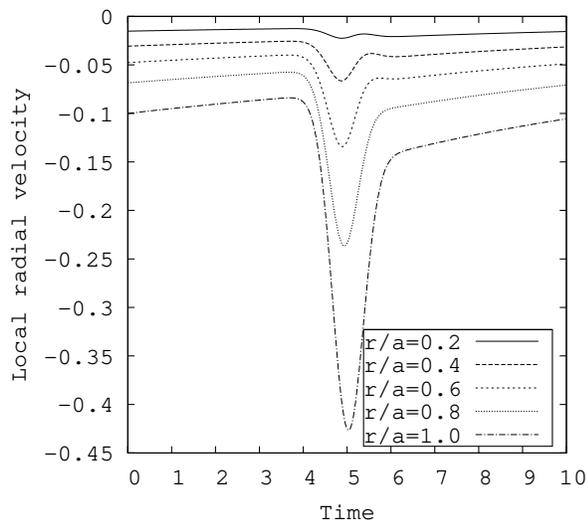}}
\caption{Evolution of the local velocity $\omega$ for the Schwarzschild--like model I.
The initial conditions are $A(0) = 5.0$, $F(0)=0.6$, $\Omega(0)=-0.1$, with $h=0.99$.}
\end{center}
\label{fig:osch}
\end{figure}
\begin{figure}[htbp!]
\begin{center}
\scalebox{0.4}{\includegraphics[angle=270]{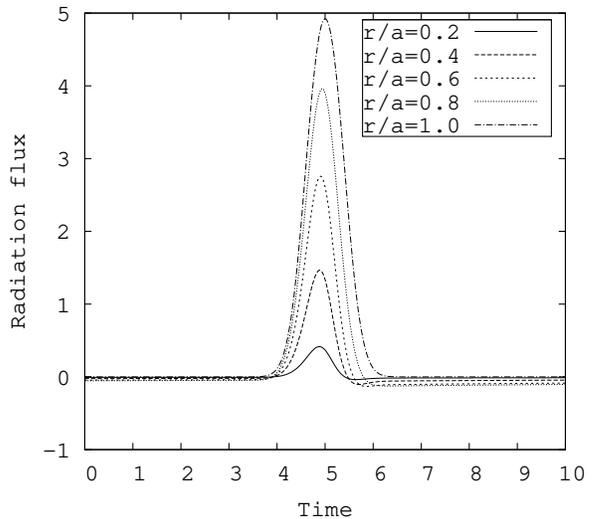}}
\caption{Evolution of the radiation flux $\epsilon$ (multiplied by $10^4$) for the Schwarzschild--like model I.
The initial conditions are $A(0) = 5.0$, $F(0)=0.6$, $\Omega(0)=-0.1$, with $h=0.99$.}
\end{center}
\label{fig:flux}
\end{figure}
\begin{figure}[htbp!]
\begin{center}
\scalebox{0.4}{\includegraphics[angle=270]{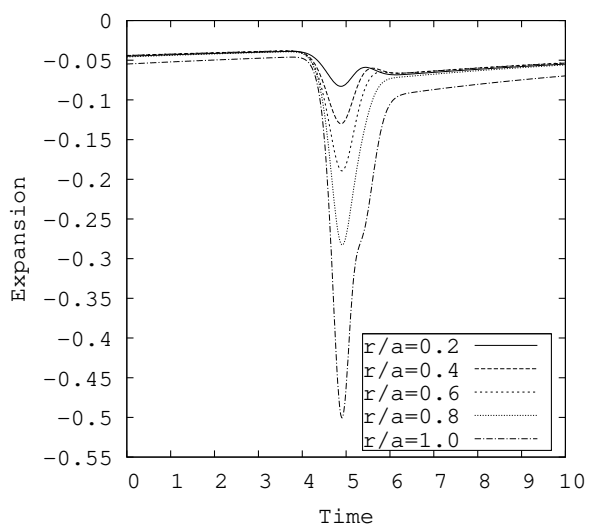}}
\caption{Evolution of the expansion $\Theta$  for the Schwarzschild--like model I.
The initial conditions are $A(0) = 5.0$, $F(0)=0.6$, $\Omega(0)=-0.1$, and $h=0.99$.}
\end{center}
\label{fig:expan}
\end{figure}
\begin{figure}[htbp!]
\begin{center}
\scalebox{0.4}{\includegraphics[angle=270]{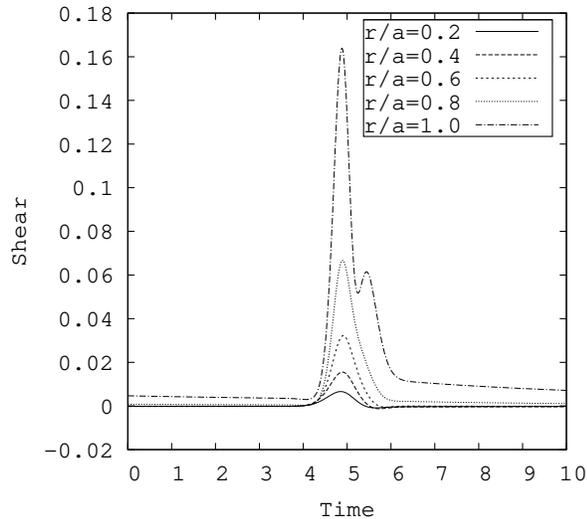}}
\caption{Evolution of the shear $\sigma$ for the Schwarzschild--like model I.
The initial conditions are $A(0) = 5.0$, $F(0)=0.6$, $\Omega(0)=-0.1$, and $h=0.99$.}
\end{center}
\label{fig:sigma}
\end{figure}
\begin{figure}[htbp!]
\begin{center}
\scalebox{0.4}{\includegraphics[angle=270]{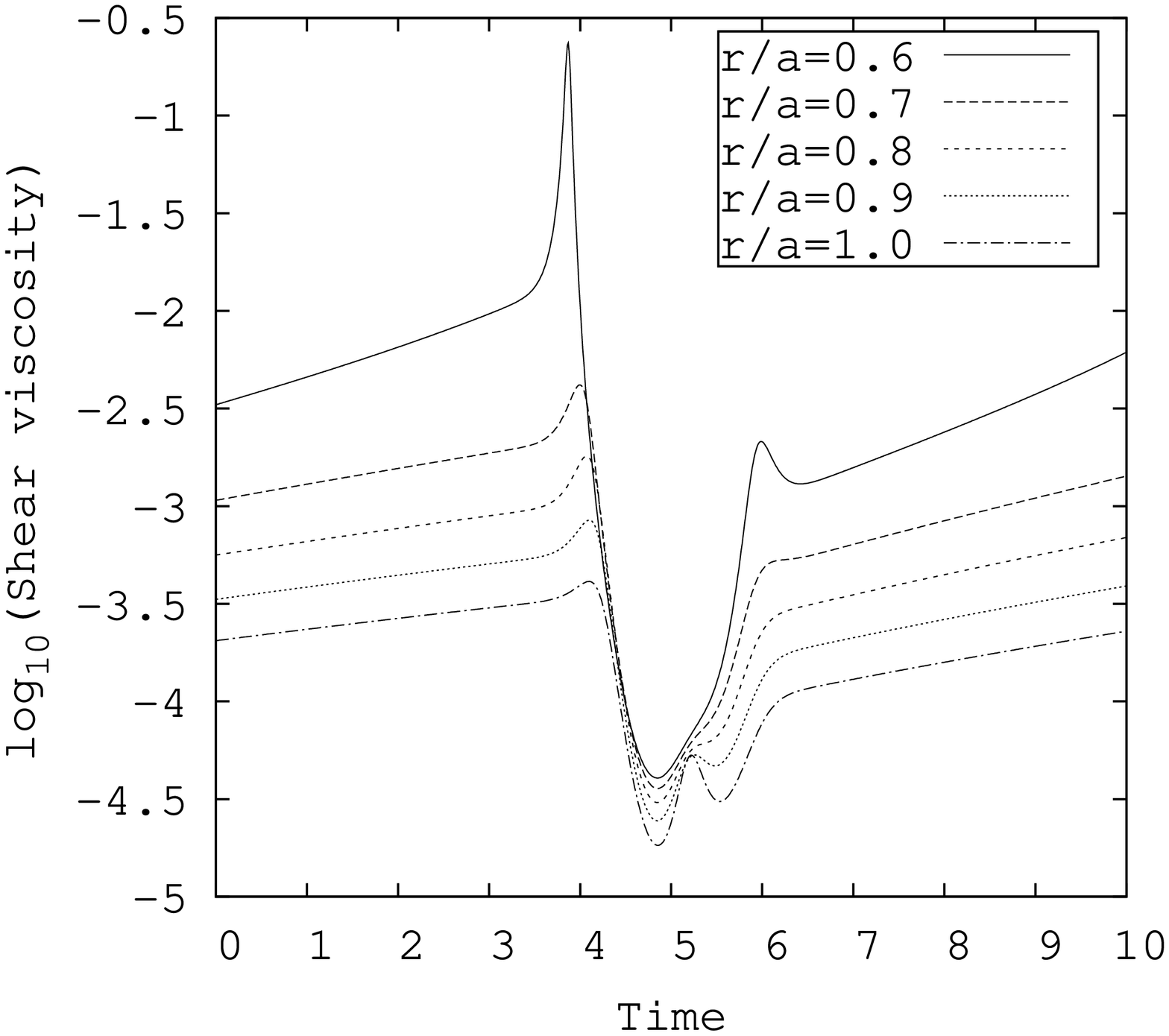}}
\caption{Evolution of the shear viscosity $\eta$ for the Schwarzschild--like model I.
The initial conditions are $A(0) = 5.0$, $F(0)=0.6$, $\Omega(0)=-0.1$, with $h=0.99$.}
\end{center}
\label{fig:visco}
\end{figure}

\section{The Postquasistatic approximation and the surface equations}
\label{sec:pqsa}

Feeding back (\ref{eq:ev1}) and (\ref{eq:ev2}) and using (\ref{eq:mass})
into (\ref{eq:ee1}) and (\ref{eq:ee2}), these two field equations may be formally 
integrated to obtain
\begin{equation}
m= \int^{r}_{0}4\pi r^2 \tilde \rho \ dr \label{eq:m}
\end{equation}
which is the Hamiltonian constraint in the ADM 3+1 formulation and
\begin{equation}
\nu=\nu_{a} + \int^r_a \frac{2(4\pi r^3 \tilde p + m)}{r(r-2m)}dr, \label{eq:nu}
\end{equation}
the polar slicing condition, from where it is obvious that for a given radial dependence of the effective
variables, the radial dependence of the metric functions becomes completely determined.

As defined in \cite{hbds2001} the postquasistatic regime is a system
out of equilibrium (or quasiequilibrium; see \cite{hm98}) but whose 
effective variables share the same radial dependence as the corresponding
physical variables in the state of equilibrium (or quasiequilibrium).
Alternatively, we can say that the system in the postquasistatic regime
is characterized by metric functions of the static (quasistatic) regime.
The rationale behind this definition is not difficult to catch: we look for
a regime which, although out of equilibrium, it is the closest to quasistatic evolution.
\subsection{The PQSA protocol}
We outline here the PQSA approach:
\begin{enumerate}

\item Take an interior solution to Einstein's field equations, representing a fluid
distribution of matter in equilibrium, with static solutions
$$
\rho_{st.}=\rho(r), \,\,\, p_{st.}=p(r).
$$

\item Assume that the $r$ dependence of $\tilde\rho$ and $\tilde p$ is the
same as that of $\rho_{st.}$ and $p_{st.}$, respectively.

\item Using equations (\ref{eq:m}) and (\ref{eq:nu}), with the $r$ dependence of
$\tilde p$ and $\tilde\rho$, one gets $m$ and $\nu$ up to some functions of
$t$, which will be specified below.
 
\item For these functions of $t$ one has three ordinary differential equations
 (hereafter referred to as the surface equations), namely:
\begin{enumerate}  
\item Equation (\ref{eq:velocity}) evaluated  at $r=a$;
    
\item Equation (\ref{eq:energy}) evaluated at $r=a$;

\item Equation $T^\nu_{1;\nu}=0$ evaluated at $r=a$.
\end{enumerate}

\item Depending on the kind of matter under consideration, the system of
     surface equations described above may be closed with the additional
     information provided by the transport equation
     and/or the equation of state for the anisotropic pressure and/or eventual
     additional information about some of the physical variables evaluated on
     the surface of the boundary (e.g. the
     luminosity).
\item Once the system of surface equations is closed, it can be integrated for
any initial data.
 
\item Feeding back the result of integration in the expressions for $m$ and
 $\nu$, these two functions are completely determined.
  
\item With the input from point 7, and using the field equations,
  together with the equation of state and/or transport equation, all
  physical variables can be found everywhere inside the
  matter distribution.
\end{enumerate}

As it should be clear from the above, the crucial point in the algorithm
is the system of equations at the surface of the distribution. We specify it in the next section.

\subsection{Surface equations}
Evaluating (\ref{eq:energy}) at the surface and using the boundary
 condition (\ref{eq:boundary}), the
 energy loss is given by   
\begin{equation}
\dot m_a =-4\pi a^2 \epsilon _a (1 - 2m_a/a) (1-\omega_a).
\end{equation}
Hereafter the overdot indicates $d/dt$ and the {\emph a} subscript 
indicates that quantity is
evaluated at the surface $r=a(t)$.

 The evolution of the boundary is governed by equation (\ref{eq:velocity})
 evaluated at the surface
\begin{equation}
\dot a=(1-2m_a/a)\omega _a.
\end{equation}
Scaling the total mass $m_a$, the radius $a$ and
 the time--like coordinate by the initial mass $m_a(t=0)\equiv m_a(0)$,
$$ A\equiv a/m_a(0), \, M\equiv m_a/m_a(0), \, t/m_a(0) \rightarrow t,$$
and defining
\begin{equation}
F\equiv 1-\frac{2M}A,
\end{equation}
\begin{equation}
\Omega \equiv \omega _a,
\end{equation}
\begin{equation}
E\equiv 4\pi a^2\epsilon _a(1-\Omega),
\end{equation}
the surface equations can be written as
\begin{equation}
\dot A=F\Omega, \label{eq:first}
\end{equation}
\begin{equation}
\dot F=\frac FA\left[(1-F)\Omega +2E\right]. \label{eq:second}
\end{equation}
Equations (\ref{eq:first}) and (\ref{eq:second}) are general
 within spherical symmetry. 
 
 We need a third surface equation to specify
 the dynamics completely for any set of initial conditions and a given
 luminosity profile $E(t)$. For this purpose we can use the field
 equation (\ref{eq:ee3}) or the conservation equation, (\ref{eq:TOV}), written in terms of the effective variables, which is clearly model--dependent.

 \begin{figure}[htbp!]
\begin{center}
\scalebox{0.4}{\includegraphics[angle=270]{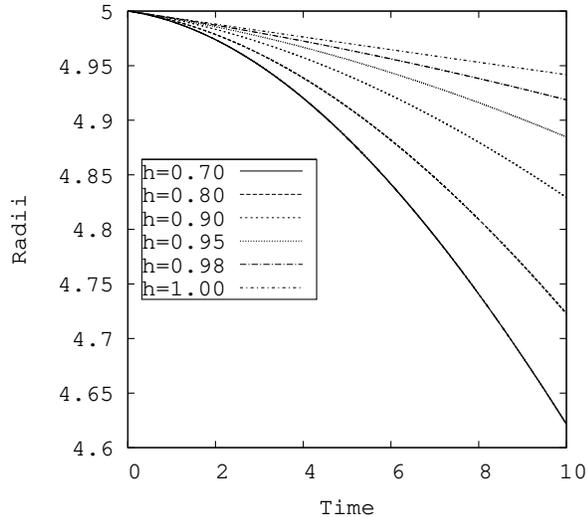}}
\caption{Evolution of the radius $A(t)$ for the Schwarzschild--like model II. The initial
conditions are $A(0) = 5$, $F(0)=0.6$, $\Omega(0)=-0.01$.}
\end{center}
\label{fig:rsch_II}
\end{figure}

\begin{figure}[htbp!]
\begin{center}
\scalebox{0.4}{\includegraphics[angle=270]{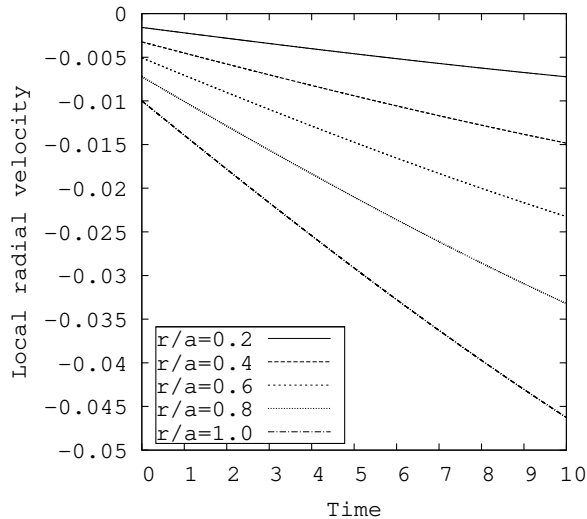}}
\caption{Evolution of the local radial velocity $\omega$  for the Schwarzschild--like model II.
The initial conditions are $A(0) = 5$, $F(0)=0.6$, $\Omega(0)=-0.01$, and $h=0.9$.}
\end{center}
\label{fig:osch_II}
\end{figure}
\begin{figure}[htbp!]
\begin{center}
\scalebox{0.4}{\includegraphics[angle=270]{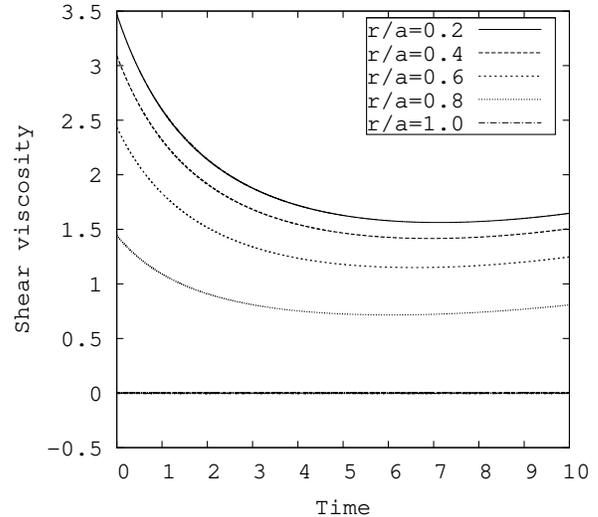}}
\caption{Evolution of the shear viscosity $\eta$ (multiplied by $10^2$) for the Schwarzschild--like model II.
The initial conditions are $A(0) = 5$, $F(0)=0.6$, $\Omega(0)=-0.01$, and $h=0.9$.}
\end{center}
\label{fig:visco_II}
\end{figure}

\section{Examples}
\label{sec:examples}
We illustrate the PQSA method with four examples based on the Schwarzschild
and Tolman VI interior solutions.
Additionally, we consider two corresponding adiabatic models, that is, without
free--streaming but anisotropic (viscous).  Although greatly simplified, the adiabatic models lead to non-trivial results, which allow to understand our results better.

\subsection{Schwarzschild--like model I: non--adiabatic}
We consider here a very simple model inspired by the well-known
Schwarzschild interior solution \cite{s16}. We take
\begin{equation}
\tilde \rho =f(t),
\end{equation}
where $f$ is an arbitrary function of $t$.
The expression for $\tilde p$ is
\begin{equation}
\frac{\tilde p + \frac{1}{3}\tilde\rho}{\tilde p + \tilde\rho}=
\left(1-\frac{8\pi}{3}\tilde\rho r^2 \right)^{h/2}k(t), \label{eq:prep}
\end{equation}
where $k$ is a function of $t$ to be defined from the boundary condition
(\ref{eq:boundary}), which now reads, in terms of the effective variables, as
\begin{equation}
\tilde p_a=\tilde\rho_a\Omega^2 + \hat{\epsilon}_a (1 + \Omega)^2.
 \label{eq:boun}
\end{equation}
Thus, (\ref{eq:prep}) and (\ref{eq:boun}) give
\begin{equation}
\tilde\rho=\frac{3(1-F)}{8\pi a^2},
\end{equation}
\begin{equation}
\tilde p=\frac{\tilde\rho}{3}\Biggl\{\frac{\chi_S F^{h/2} -3\psi_S\xi}
{\psi_S\xi -\chi_S F^{h/2}}\Biggr\}, \label{eq:effepre}
\end{equation}
with
$$
\xi=[1-(1-F)(r/a)^2]^{h/2}
$$
where $h=1-2C$ and
$$\chi_S=3(\Omega^2+1)(1-F)+2E(1+\Omega),$$

$$\psi_S=(3\Omega^2+1)(1-F)+2E(1+\Omega).$$
Using (\ref{eq:mass}) and (\ref{eq:nu}) it is easy to obtain expressions
for $m$ and $\nu$:
\begin{equation}
m=m_a(r/a)^3, \label{eq:mass_sch}
\end{equation}
\begin{equation}
e^{\nu}=\Biggl\{\frac{\chi_S F^{h/2}-\psi_S\xi}{2(1-F)}
\Biggr\}^{2/h}. \label{eq:nu_sch}
\end{equation}
In order to write down explicitely the surface equations for this example,
we evaluate the equation (\ref{eq:TOV}) at the surface, obtaining
\begin{eqnarray}
\dot{\Omega}&=&[8EF-\Omega^2+10\Omega^2F-6E\Omega+2E\Omega^2+3\Omega^4-8E^2   \nonumber\\
&-&9\Omega^2F^2-6\Omega^4F +8E\Omega^3+3F^2\Omega^4+4E^2\Omega+4E^2\Omega^2 \nonumber\\
&+& 4\dot{E}A+6FE\Omega-2F\Omega^2E-8F\Omega^3E) \nonumber \\
&/&(2A(F-1))
\label{eq:opunto1}
\end{eqnarray}

It is interesting to note that this equation is the same as in the
isotropic case ($p_r = p_t$). This is a direct consequence of the
chosen equation of state combined with incompressibility of the fluid;
it is not a general result, as we will see for the next models.
Equation (\ref{eq:opunto1}), together with (\ref{eq:first}) and (\ref{eq:second}),
 constitute the system of differential equations at the surface for this model.  It is
necessary to specify one the luminosity as a function of $t$ and the initial data.
We choose $E$ to be a gaussian 
$$E=E_0 e^{-(t-t_0)^2/\Sigma^2},$$
with $E_0=M_r/\sqrt{\Sigma \pi}$, $t_0=5.0$ and $\Sigma=0.25$, which corresponds to a pulse
radiating away $M_r = 1/10$ of the initial mass. 

We solve equations (\ref{eq:first}), (\ref{eq:second}), and 
(\ref{eq:opunto1}) using a fourth order
Runge-Kutta method. The physical variables ($\rho$, $p$, 
$\omega$, $\eta$, $\epsilon$)
are obtained from the field equations (\ref{eq:ee1})--(\ref{eq:ee4}) and the equation of state (\ref{eq:EDE}). Note that we have to use equations (\ref{eq:expansion}) and (\ref{eq:shear}) and some additional numerical work to determine $\eta$. We take as initial conditions $A(0)=5$,
$M(0)=1$, $\Omega(0)=-0.1$, with $h=0.99$.

Figure 1 shows the evolution of the radius of the distribution. Figures 2--6 display the physical variables ($\rho$, $p_r$, $p_t$, $\omega$, $\epsilon$), figures 7--8 the kinematic variables $\Theta$ and $\sigma$, and  figure 9 the shear viscosity $\eta$, for different regions.
It is evident that the emission of energy decreases the energy density and 
the shear viscosity, but increases the pressure; while the collapse is briefly accelerated.
It is interesting to note that after the
gaussian emission the distribution recovers staticity slowly, probably in a quasistatic regime.
In this model $p_t > p_r$, which means $2\sqrt 3 \sigma \eta > 0$ ($h=0.99$). 
It is important to mention that in this model, a shear viscosity $\eta >0$ is only possible if we choose the negative root in (\ref{eq:shear}). 
Physically meaningful values of shear viscosity ($\eta>0$) are obtained for regions $r/a \approx 0.6\rightarrow 1$. This means that the inner core is not
viscous but anisotropic. The rest of the kinematic variables
($\Theta$ and $\sigma$), shown in Figures 7--8, follow the evolution 
of the radius of the distribution, with $\Theta$ ($\sigma$) decreasing (increasing) faster as the radius decreases at a faster rate for $4 \lsim t \lsim 6$.

\begin{figure}[htbp]
\begin{center}
\scalebox{0.4}{\includegraphics[angle=270]{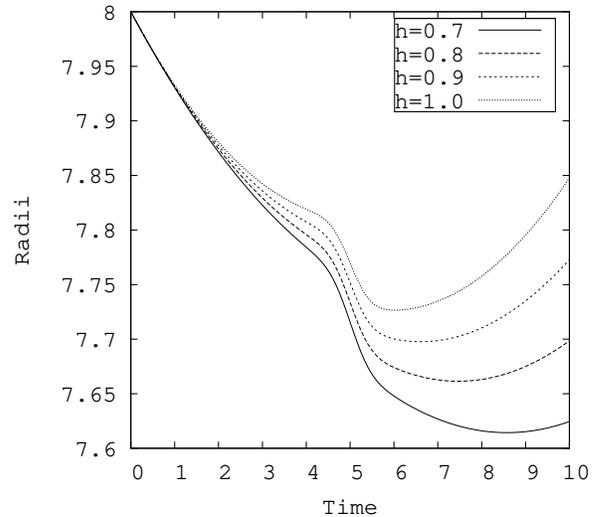}}
\caption{Evolution of the radii for the Tolman VI-like model III. The initial
conditions are $A(0) = 8.0$, $F(0)=0.75$, $\Omega(0)=-0.1$.}
\end{center}
\label{fig:rt6}
\end{figure}

\begin{figure}[htbp!]
\begin{center}
\scalebox{0.4}{\includegraphics[angle=270]{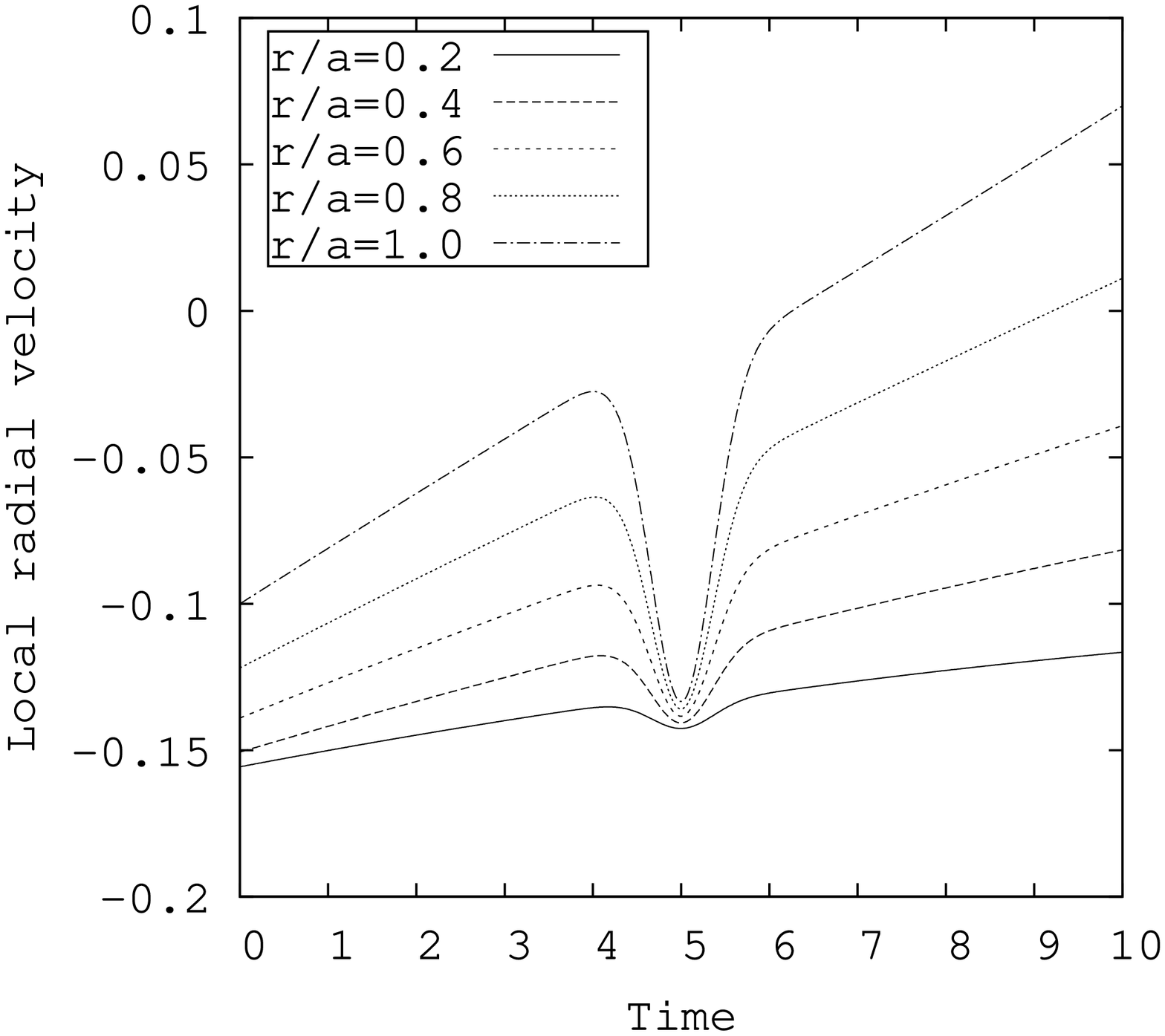}}
\caption{Evolution of the local radial velocity $\omega$  for the Tolman VI-like model III
The initial conditions are $A(0) = 8.0$, $F(0)=0.75$, $\Omega(0)=-0.1$, with $h=0.95$.}
\end{center}
\label{fig:ot6}
\end{figure}
\begin{figure}[htbp!]
\begin{center}
\scalebox{0.4}{\includegraphics[angle=270]{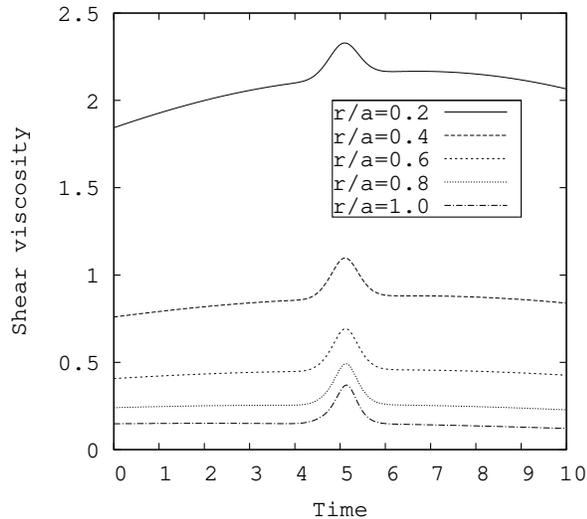}}
\caption{Evolution of the shear viscosity $\eta$ (multiplied by $10^4$) for the Tolman VI-like model III.
The initial conditions are $A(0) = 8$, $F(0)=0.75$, $\Omega(0)=-0.1$, with $h=0.95$.}
\end{center}
\label{fig:eta_t6}
\end{figure}

\subsection{Schwarzschild--like model II: adiabatic}
We construct this model with the same effective variables
and metric functions as the aforestudied model I, but now the radiation flux is zero
everyhere; therefore this model is adiabatic. Obviously we do not need now
an equation of state because all physical variables are determined
algebraically from the field equations. However some measure of
tangential stress at the surface is required to evolve the system. We
opt for a tangential pressure equal to the radial pressure just
at the surface, $p_t|_a=p_r|_a$. The third surface equation in this case
is 
\begin{eqnarray}
\dot\Omega&=&-\frac{1}{2A}(4h\Omega^2F-\Omega^2F+hF-6\Omega^4F+3h\Omega^4F\nonumber\\
&-&F-3h\Omega^4-4h\Omega^2-3\Omega^2+1-h+6\Omega^4)
\end{eqnarray}
Observe that this expression explicitly depends on the anisotropic
parameter $h$. In this case we integrate the system for the initial
conditions $A(0)=5$, $M(0)=1$ and $\Omega=-0.01$.
Figure 10 shows the radius of the distribution for different values of $h$. Figures 11--12 display
the radial velocity and the shear viscosity for $h=0.9$.
In this case anisotropy manifests clearly at the surface. As long as $p_t$ is greater than $p_r$ the collapse accelerates. The same occurs for $0.7 \leq h \leq 1.0$, as we go deeper in the distribution the inner shells collapse faster. The fffective gravitation is therefore enhanced by the anisotropy induced by the viscosity. Inner regions have a greater shear viscosity in this model ($\sim 10$-$10^5$ times the values found in model I).

\subsection{Tolman VI--like model III: non--adiabatic}
In this subsection we revise the model obtained from Tolman's solution
VI \cite{t39}. Let us take
\begin{equation}
\bar\rho=\frac{g}{r^2},
\end{equation}

\begin{equation}
\tilde p = \frac{gK(1-9 \alpha (r/a)^{\sqrt{4-3h}})}{3(K/I-9 \alpha (r/a)^{\sqrt{4-3h}}) h r^2},
\end{equation}
where $K$ and $I$ are defined as
$$
K=8-3h+4\sqrt{4-3h},
$$
$$
I=8-3h-4\sqrt{4-3h}.
$$
$g$  and $\alpha$ are functions of $t$, which can be determined using
(\ref{eq:boun}). Therefore

\begin{equation}
g=\frac{3(1-F)}{24\pi } \label{eq:g}
\end{equation}
\begin{equation}
\alpha=\frac{3h(1-F)-K\beta}{9[3h(1-F)-I\beta]}
\end{equation}
where
$$\beta= (1-F)\Omega^2+2E(1+\Omega).$$
Using (\ref{eq:mass}) and (\ref{eq:nu}) we obtain
\begin{equation}
m=m_{a}(r/a)
\end{equation}
and
\begin{widetext}
\begin{eqnarray}
{\nu} & = & \ln F + \frac{8 \pi g}{F} \left(1+\frac{I}{3h}\right)\ln(r/a)
  +  \frac{8\pi g}{3 h F \sqrt{4-3h}}\left\{ I \ln\left(\frac{ (K/I-9\alpha)a^{\sqrt{4-3h}} }{ (K/I) a^{\sqrt{4-3h}}-9\alpha r^{\sqrt{4-3h}}} \right) \right. \nonumber \\
& & \left.  +  K \ln\left(\frac{ (K/I)a^{\sqrt{4-3h}}-9\alpha r^{\sqrt{4-3h}}}{a^{\sqrt{4-3h}}(K/I-9\alpha)}\right)   \right\}
\end{eqnarray}
Evaluating equation (\ref{eq:TOV}) at the surface we can obtain an equation for
$\dot{\Omega}$ (too long to display here).

Integrating the system of equations at the surface for the initial conditions 
$A(0)=8$ and $\Omega=-0.1$, with $M_r=10^{-2}$, we obtain figures 13, 14 and 15. We obtain similar results as in model II but in a different fashion. The bigger the difference between
the tangential (viscous) pressure $p_t$ and the radial pressure $p_r$ (as $h$ decreases), the more violently the distribution explodes.
It is striking that now {\it all} the spherical shells tend to reach the same instantaneous
local radial velocity when the system goes to faster collapse with emission of energy across
de boundary surface. At least locally, the ``acceleration" of all the shells goes
to zero at the same time; again the same instantaneous local radial velocity (negative)
is reached before a final bouncing per shell from outer to inner.
 
\subsection{Tolman VI--like model IV: adiabatic}
We construct this model with the same effective variables
and metric functions as in model III, but now the radiation flux is zero
everyhere; therefore this model is adiabatic. Obviously we do not need now
an equation of state because all physical variables are determined
algebraically from the field equations. However some measure of
tangential stress at the surface is required to evolve the system. We
opt for a tangential pressure equal to the radial pressure just
at the surface, $p_t|_a=p_r|_a$, as in model II. The third surface equation in this case
is

\begin{eqnarray}
\dot\Omega&=&(-9\Omega^4Kh^2I^2F+\Omega^4FI^2K^3+9\Omega^4K^2h^2IF+
             6\Omega^4FI^2K^2 h\sqrt{4-3h}-243FIh^4\Omega^2-I^3K^2F\Omega^4\nonumber\\
     &-&81Fh^4I+54FIKh^3\sqrt{4-3h}+243Kh^4\Omega^2F-18\Omega^2FI^2Kh^2\sqrt{4-3h}
     -18\Omega^2FIK^2h^2\sqrt{4-3h}\nonumber \\
     &+&81Fh^4K
     -18h^2I^2FK\Omega^2+18Fh^2K^2\Omega^2I+9\Omega^4Kh^2I^2
     -9\Omega^4K^2h^2I-18h^2K^2\Omega^2I +18h^2I^2K\Omega^2\nonumber\\
     &+&I^3K^2\Omega^4-81h^4K-I^2\Omega^4K^3+81h^4I+81Kh^4\Omega^2 
     -81Ih^4\Omega^2)/[162(K-I)h^4A]
\end{eqnarray}
\end{widetext}
Integrating the system for the initial
conditions $A(0)=8$ and $\Omega=-0.1$, 
figure 13 shows the radius of the distribution for different values of $h$. Figures 16--18 display
the radial velocity and the shear viscosity for $h=0.9$. After some numerical experimentation some
non--trivial results arise, and we relax the condition $p_t>p_r$. 
At the surface we do not find any novelty. 
The most violent explosion occurs as $p_r >> p_t$.  In this adiabatic
but viscous (anisotropic) model all the shells bounce at the same time to irrupt from inner
regions to outer regions with an apparently linear dependence with time. 
The outer shells of matter are ejected faster and earlier than the inner
ones. This sort of behavior was reported several years ago 
studying in Bondi coordinates the collapse of radiating distributions with an 
extreme transport mechanism as diffusion \cite{bhs89}. 
However, the shear viscosity profiles indicate that i) bouncing is not allowed at all and ii) some inner regions are forbidden, otherwise the shear viscosity profiles become negative or/and infinite (see Figure 18). 
This situation is general and independent of the anisotropy parameter $h$. 

\begin{figure}[htbp]
\begin{center}
\scalebox{0.4}{\includegraphics[angle=270]{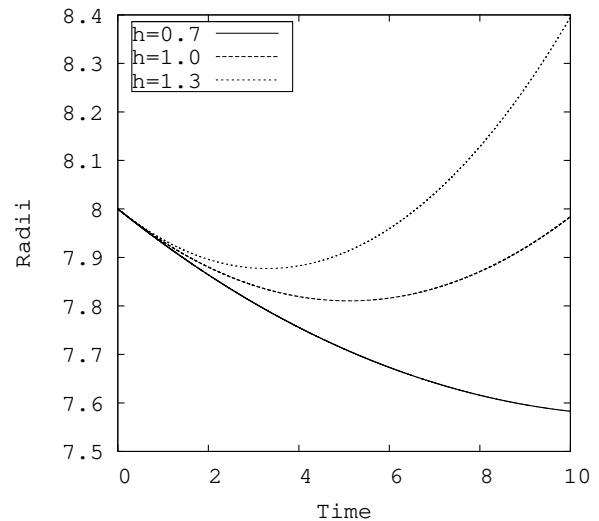}}
\caption{Evolution of the radii for the Tolman VI-like model IV. The initial
conditions are $A(0) = 8.0$, $F(0)=0.75$, $\Omega(0)=-0.1$.}
\end{center}
\label{fig:rtolsixa}
\end{figure}
\begin{figure}[htbp!]
\begin{center}
\scalebox{0.4}{\includegraphics[angle=270]{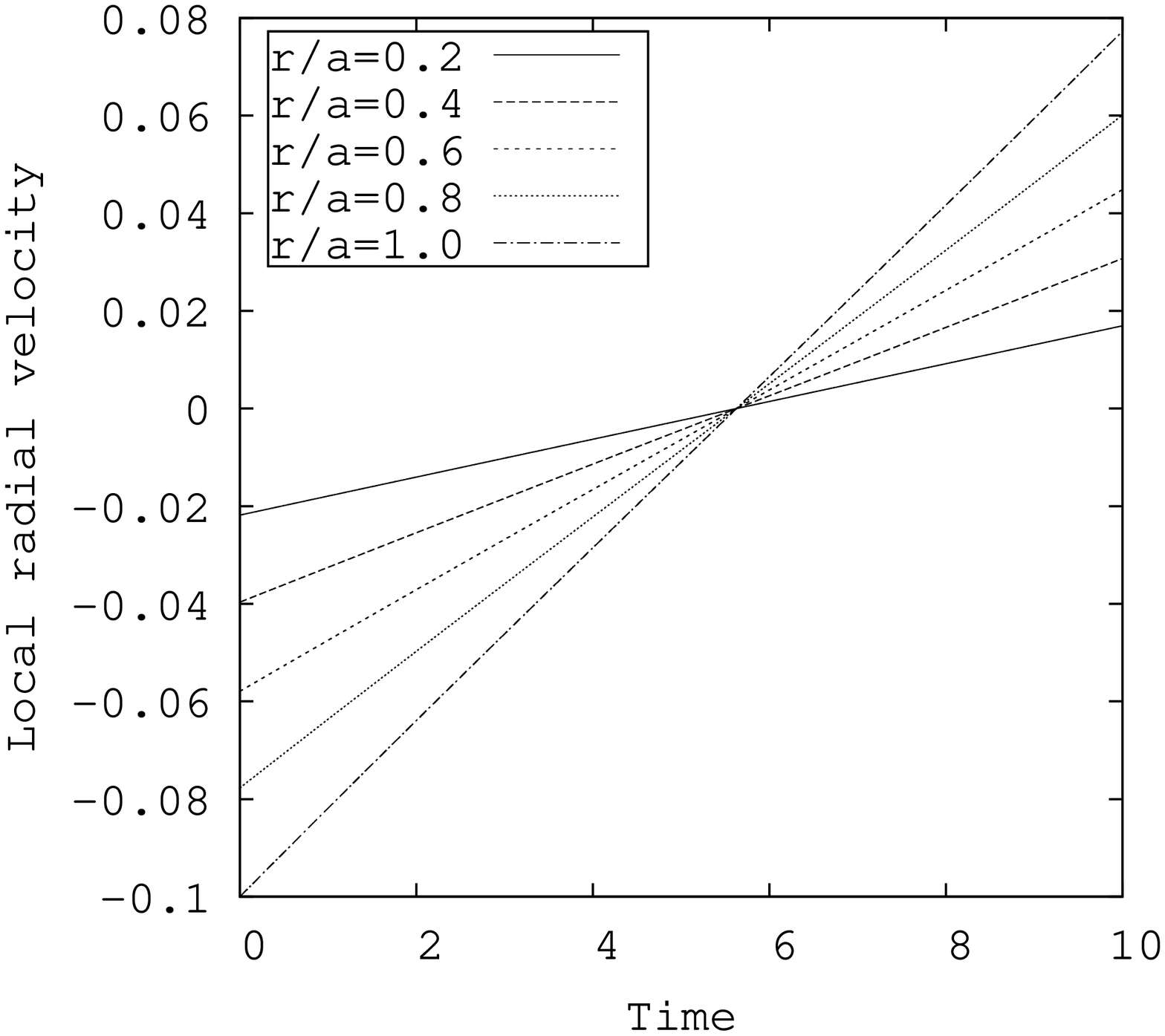}}
\caption{Evolution of the local radial velocity $\omega$  for the Tolman VI-like model IV.
The initial conditions are $A(0) = 8.0$, $F(0)=0.75$, $\Omega(0)=-0.1$, with $h=0.95$.}
\end{center}
\label{fig:otolsixa}
\end{figure}
\begin{figure}[htbp!]
\begin{center}
\scalebox{0.4}{\includegraphics[angle=270]{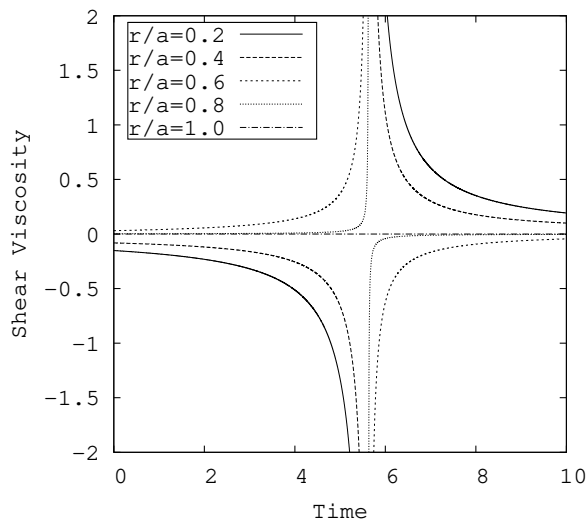}}
\caption{Evolution of the shear viscosity $\eta$ for the Tolman VI-like model IV.
The initial conditions are $A(0) = 8$, $F(0)=0.75$, $\Omega(0)=-0.1$, with $h=0.95$.}
\end{center}
\label{fig:etatosixa}
\end{figure}
\section{Conclusions}
\label{sec:conc}
We consider a selfgravitating
spherical distribution of matter containing a dissipative fluid. 
The use of the PQSA with non--comoving coordinates
allow us to study viscous fluid spheres in the streaming out limit as they just depart from equilibrium. From this point of view,
the PQSA can also be seen as a nonlinear perturbative method 
to test the stability of solutions in equilibrium.

For the non--adiabatic Schwarzschild model the distribution
evolves to a final state with a non--viscous and 
anisotropic inner core. Surprisingly, in this model the evolution
of the local radial velocity at the surface is the same
in the isotropic ($p_t=p_r$) case, a fortuitous coincidence
due to the chosen equation and state and the incompressibility of the fluid. 
For the adiabatic Schwarzschild model the final
core is up to $10^5$ times more viscous, and the anisotropy
appears explicitly in all the evolution equations. The higher
viscosity of the core increases the effective gravity and
the collapse is faster, as long as $p_t > p_r$.

Both of the Tolman VI models lead to a distribution which initially
collapses and then bounces and expands indefinitely.
The Tolman VI non--adiabatic model shares some of the characteristics of the adiabatic Schwarzschild.
Before the final bouncing, as $p_t > p_r$ the collapse is accelerated.
For the non--adiabatic case some regions of the parameter space are
forbidden, since the shear viscosity profiles become unphysical. In this
case the  bouncing is not allowed and the distribution collapses indefinitely.

A forthcoming paper considers the dissipation by
heat flow, in order  to isolate effects similar to the
ones studied in the present investigation, but with different
mechanisms. Also, a work in progress considers heat flow
and anisotropy induced by electric charge,  
pointing to the most realistic numeric modeling in
this area  \cite{dhlms07}. 
Although they are not entirely new, the results
presented here constitute a first cut to more general situations
using the PQSA, including dissipation, anisotropy, electric charge, heat flow,
viscosity, radiation flux, superficial tension, temperature profiles
and study their influence on the gravitational collapse.
This investigation is an essential part of a long-term project which 
tries to incorporate the M\"uller--Israel--Stewart theory for dissipation and deviations from spherical symmetry, specially when considering electrically charged distributions. Besides being interesting in their own right, we believe that spherically symmetric fluid models are useful as a test bed for more general solvers 
in numerical relativity \cite{bcb09,bcb10}. A general 3D code must be able
to reproduce situations closer to equilibrium.

\begin{acknowledgments}
WB was on sabbatical leave from Universidad de Los Andes
while finishing this work. CP acknowledges the computing resources provided
by the Victorian Partnership for Advanced Computation (VPAC).
\end{acknowledgments}
\bibliographystyle{apsrev}
\bibliography{visco}
\end{document}